\documentclass[twocolumn,trackchanges,twocolappendix]{aastex63}
\usepackage{xcolor}

\DeclareRobustCommand{\ers}{\bgroup\markoverwith{\textcolor{red}{\rule[.5ex]{2pt}{0.4pt}}}\ULon}

\received{}
\revised{}
\accepted{}
\submitjournal{ApJ}
\shorttitle{Neutrino Constraints on Radio Emission from Massive Clusters}
\shortauthors{Nishiwaki, Asano \& Murase}
\begin{document}

\title{
High-Energy Neutrino Constraints on Cosmic-ray Re-acceleration \\in Radio Halos of Massive Galaxy Clusters
}

\author{Kosuke Nishiwaki}
\affiliation{Institute for Cosmic Ray Research, The University of Tokyo, 5-1-5 Kashiwanoha, Kashiwa, Chiba 277-8582, Japan}

\author{Katsuaki Asano}
\affiliation{Institute for Cosmic Ray Research, The University of Tokyo, 5-1-5 Kashiwanoha, Kashiwa, Chiba 277-8582, Japan}

\author{Kohta Murase}
\affiliation{Department of Physics, The Pennsylvania State University, University Park, Pennsylvania 16802, USA}
\affiliation{Department of Astronomy \& Astrophysics, The Pennsylvania State University, University Park, Pennsylvania 16802, USA}
\affiliation{Center for Multimessenger Astrophysics, Institute for Gravitation and the Cosmos, The Pennsylvania State University, University Park, Pennsylvania 16802, USA}
\affiliation{Institute for Advanced Study, Princeton, New Jersey 08540, USA}
\affiliation{Yukawa Institute for Theoretical Physics, Kyoto University, Kyoto 606-8502, Japan}

\begin{abstract}
A fraction of merging galaxy clusters host diffuse radio emission in their central region, termed as a giant radio halo (GRH).
The most promising mechanism of GRHs is the re-acceleration of non-thermal electrons and positrons by merger-induced turbulence. 
However, the origin of these seed leptons has been under debate, and either protons or electrons can be primarily-accelerated particles. 
In this work, we demonstrate that neutrinos can be used as a probe of physical processes in galaxy clusters, and discuss possible constraints on the amount of relativistic protons in the intra-cluster medium with the existing upper limits by IceCube. We calculate radio and neutrino emission from massive ($>10^{14}M_\odot$) galaxy clusters, using the cluster population model of \citet{paperII}. This model is compatible with the observed statistics of GRHs, and we find that the contribution of GRHs to the isotropic radio background observed with the ARCADE-2 experiment should be subdominant. Our fiducial model predicts the all-sky neutrino flux that is consistent with IceCube's upper limit from the stacking analysis. We also show that the neutrino upper limit gives meaningful constraints on the parameter space of the re-acceleration model, such as electron-to-proton ratio of primary cosmic-rays and the magnetic field, and in particular the secondary scenario, where the seed electrons mostly originate from inelastic $pp$ collisions, can be constrained even in the presence of re-acceleration.
\end{abstract}

\section{Introduction\label{sec:intro}}
Galaxy clusters (GCs) have drawn attention as possible sources of high-energy neutrinos observed in IceCube \citep[][]{IceCube_2013PhRvL.111b1103A,IceCube:2013low}. 
They can work as reservoirs of cosmic rays (CRs), within which cosmic-ray protons (CRPs) are accumulated over a cosmological timescale and emit gamma rays and neutrinos through the inelastic $pp$ collisions with the cold protons in the intra-cluster medium (ICM) \citep[e.g.,][]{Murase:2013rfa}. 
\par

Several previous studies compared two possibilities for the source of CRs in the ICM; the ``accretion shock" model and the ``internal source" model \citep[e.g.,][]{Murase_Waxman2016}. 
In the accretion shock model, CRs are accelerated at the shocks formed around the virial radius of the clusters, and at most $\sim20\%$ of the IceCube flux around 1 PeV can be explained by the emission from the GCs with $M>10^{15}M_\odot$ \citep[e.g.,][]{Zandanel2015,Fang_Olinto2016,Hussain:2021dqp}. 
In the internal source models, the CRs are provided through the outflows from active galactic nuclei (AGN) and galaxies inside the cluster. 
In this case, previous studies showed that nearly $100\%$ of the IceCube flux can be achievable \citep[e.g.,][]{Murase2008,Kotera:2009ms,Fang_2018NatPh..14..396F,Hussain:2021dqp}. 
\par

Recently, \citet{IceCube_2022_stack} reported an upper limit on the contribution of massive galaxy clusters to the diffuse muon-neutrino flux derived from a stacking analysis of 1,094 clusters in {\it Planck} Sunyaev-Zel'dovich (SZ) sources.
The contribution from the clusters with masses of $10^{14}M_\odot < M_{500} < 10^{15}M_\odot$ in a redshift $0.01 < z < 2$ is limited to be less than $\sim$5\% at 100 TeV, although the limit strongly depends on the weighting scheme to account for the completeness of the {\it Planck} catalog. 
Some of the accretion shock models \citep[e.g.,][]{Hussain:2021dqp} are in tension with this limit, while the internal source models, where the background is dominated by lower mass clusters, are not constrained. This result is complementary to neutrino anisotropy limits~\cite{Murase_Waxman2016}, which also disfavor the accretion shock models for the all-sky neutrinos at PeV energies.
\par

On the other hand, direct evidence of relativistic electrons in the ICM has been achieved with radio observations of synchrotron emission.
A fraction of clusters host extended radio structures, known as giant radio halos (GRHs), mini halos, and cluster radio shocks \citep[][]{review_vanWeeren}. 
The most plausible mechanism of GRHs is so-called turbulent re-acceleration, where the non-thermal electrons are re-energized by the stochastic interaction with the merger-induced turbulence permeating the cluster volume \citep[see][for a review]{Brunetti_Jones_review}. 
\par

Inelastic $pp$ collisions between CRPs and protons in the ICM can provide relativistic electrons through the decay of $\pi$ mesons. 
However, there is a growing number of pieces of evidence against the purely hadronic origin of radio-emitting CREs. 
For example, a number of GRHs and diffuse emissions in the cluster periphery are found with a very steep spectral index ($F_\nu\propto \nu^{-\alpha_{\rm syn}}$ with $\alpha_{\rm syn}\gtrsim 1.5$) \citep[e.g.,][]{ Macario_2013A&A...551A.141M, Wilber_2018MNRAS.473.3536W, Duchesne_2021PASA...38...53D,Bruno_2021A&A...650A..44B,Botteon_2022}. 
Such steep spectra challenge the pure hadronic model, since the energy budget required for non-thermal components including CRPs, CREs, and the magnetic field exceeds the thermal energy density of the ICM \citep[e.g.,][]{Brunetti_2008Natur.455..944B}.
The non-detection of gamma-ray from nearby GCs is in tension with the pure hadronic model, unless the magnetic field is much stronger than 1~$\mu$G \citep[e.g.,][]{Jeltema_2011, Ackermann2016_Coma, Brunetti2017}. 
In addition, the clear bi-modality between GRHs and clusters without diffuse emission supports the re-acceleration model rather than the pure hadronic model \citep[e.g.,][]{Cassano_2012,2021A&A...647A..51C,Cassano_2023}.
\par

While the pure hadronic model of GRHs is facing various problems, the $pp$ collision of CRPs can be a viable mechanism for providing ``seed" electrons for the re-acceleration.
We refer to this as the ``secondary scenario" for the seed population.
This scenario requires less amount of CRPs than the pure hadronic model and can explain the observed spectrum and profile of the GRH in the Coma cluster without violating the gamma-ray upper limit \citep[e.g.,][]{Brunetti2017,Pinzke2017,paperI}. 
As discussed in \citet{Brunetti2017}, the gamma-ray limit of Coma can be used to constrain the parameters in this secondary re-acceleration model (see Sect.~\ref{sec:constraint}).

\par

Non-thermal emission from GCs may be related to the radio background measured with the Absolute Radiometer for Cosmology, Astrophysics, and Diffuse Emission (ARCADE) 2 experiment \citep[][]{ARCADE2_Fixsen_2011ApJ...734....5F}.
The excess emission over the cosmic microwave background radiation (CMB) and the Galactic foreground is often termed as ``ARCADE-2 excess", and the origin of this excess is still under debate.
\citet{Fang_2015PhRvD..91h3501F} pointed out that the radio emission from GCs could be the origin of the ARCADE-2 excess.
\par


In this paper, we discuss the possible constraints on the secondary scenario for GRHs from the neutrino stacking analysis on galaxy clusters \citep{IceCube_2022_stack}.
This can be done by calculating the neutrino flux from the population of GCs using the re-acceleration model consistent with various observational properties of GRHs.
The same method allows us to revisit the radio background from GCs, which could contribute to the ARCADE-2 excess.
\par

In our previous study \citep[][]{paperII}, the population of GRHs is discussed using the merger tree of dark matter halos built with a Monte Carlo method.
The model parameters for the re-acceleration model are constrained by the statistical properties of GRHs at 1.4 GHz, such as the fraction of GRHs $f_{\rm RH}$ \citep[e.g.,][]{Cassano2010,Cuciti_2015A&A...580A..97C}.
Solving the Fokker--Planck (FP) equation for various realization of cluster parameters, the method in \citet[][]{paperII} can study how the efficiency of the re-acceleration depends on the cluster properties, such as the mass and the gas temperature.
Thus, our method is ideal for the estimate of the background contributions of GCs in consideration of both the physics of the re-acceleration and the constraints from the GRH observations.
\par

This paper is organized as follows. 
In Sect.~\ref{sec:LF}, we review our method to construct the luminosity functions (LFs) of non-thermal emissions from GCs, using the merger tree of dark matter halos and the luminosity--mass (LM) relation.
In Sect.~\ref{sec:contribution_backgound}, we show that the contribution of GCs to the ARCADE-2 excess is sub-dominant. 
The contribution to the IceCube flux around 100 TeV is as large as $\approx5\%$, which is comparable to the upper limit placed with the stacking analysis.
In Sect.~\ref{sec:constraint}, we discuss the constraints on the parameters of the re-acceleration model from the neutrino upper limit.
In Sect.~\ref{sec:discussion}, we 
summarize our results and discuss the possible constraints with future radio and neutrino observations.
\par

Throughout this study, we assume the secondary scenario for the seed CREs to obtain the theoretically largest intensity of the high-energy neutrino emission from GCs.
It has already been shown that this scenario is compatible with the observed population of GRHs \citep[][]{paperII}.
We adopt the flat $\Lambda$CDM cosmology with the parameters from \citet{Planck18}, where $H_0 = 67~{\rm km/s/Mpc}$ ($h = 0.67$ or $h_{70} = 0.96$). 
\par

\section{Merger Tree and Luminosity Function}\label{sec:LF}
Following \citet{paperII}, we construct the LFs of radio and neutrino emissions from GCs, combining the merger tree of GCs and the LM relations, which are obtained from the Monte Carlo method and solving spatially 1D FP equations, respectively.
Our method incorporates the re-acceleration model, and the LM relations depend on the merger history (Section~\ref{subsec:L_evolution}).
Most importantly, our model agrees with the observed statistical properties of GRHs.
\par 

Here, we briefly review our method \citep[see][for the details]{paperII}.
We follow the merger history for GCs, assuming that major mergers ignite the re-acceleration of CRs.
The luminosity depends on both the cluster mass $M_{500}$ and the time since the onset of the re-acceleration $t_{\rm R}$. 
The time evolution of the luminosities is studied by calculating the FP equations for CRPs and CREs, where we take into account radiative and Coulomb energy losses, re-acceleration, primary CRP injection and secondary CRE production by $pp$ collision, and loss of CRPs due to the $pp$ collisions.
The primary CRP is injected with a single power-law spectrum with the spectral index of $\alpha_{\rm inj}$.
\par

For the magnetic field, we assume the scaling $B(r) = B_0(\frac{n_{\rm ICM}(r)}{n_{\rm ICM}(0)})^{\eta_B}$, where $n_{\rm ICM}(r)$ is the ICM number density as a function of radius $r$.
For $\eta_B$ and $B_0$, we adopt the best-fit values $(B_0,\eta_B) = (4.7\mu{\rm G}, \eta_B)$ obtained for the Coma cluster with the measurement of the Faraday rotation measure (RM) \citep[][]{Bonafede2010}. We adopt the ICM density profile of the Coma cluster \citep[][]{Briel_1992A&A...259L..31B}.

The parameters for the primary CRP injection and the turbulent acceleration are adjusted to reproduce the observed spectrum and the brightness profile of the GRH in the Coma cluster \citep[][]{paperI}.
In our secondary model, the CRP energy density of the Coma cluster is estimated to be $\epsilon_{\rm CRP}\approx 3\times10^{-13}~{\rm erg/cm^3}$.
As discussed in Sect.2.5 of \citet{paperII}, this model is consistent with the gamma-ray upper limit on the Coma cluster given by the Fermi observation \citep[][]{Ackermann2016_Coma}.
The neutrino emissivity is calculated consistently with the radio emissivity. 

\par

Our model has three parameters concerning the radio luminosity: the threshold for the merger mass ratio, $\xi_{\rm RH}$, above which the merger can ignite the re-acceleration of CRs, and the slope and the normalization of the LM relation.
We adopt the same values for those parameters as \citet{paperII}, which can reproduce the number count of the RHs observed at 1.4 GHz (see Fig.7 of \citet{paperII}).
We have other two parameters for the neutrino LM relation.
Those two parameters can be obtained from the calculations in our re-acceleration model by introducing a mass dependence
of the momentum diffusion coefficient for CRs, $D_{pp}$ (see Sect.~\ref{subsec:model_parameters}). The mass dependence in the re-acceleration  efficiency is suggested from the radio LM relation \citep[e.g.,][]{Cassano_2012,paperII}.
\par

Since the LM relations have crucial effects on the flux of the background emission, we test two models; the fiducial model and the optimistic model.
The latter model results in larger background intensities, but the optimistic LM relation is in tension with the RH observations \citep[e.g.,][]{2021A&A...647A..51C,Cuciti_2023}.
\par


\subsection{Luminosity Evolution in Re-acceleration Model}
\label{subsec:L_evolution}
\par

We adopt the following form for the LM relations for radio and neutrino emissions, 
\begin{eqnarray}
\left(\frac{L(M_{500},t_{\rm R})}{L_0}\right) = 10^{a}\left(\frac{M_{500}}{M_0}\right)^{b}g(t_{\rm R}),
\label{eq:Luminosity_model}
\end{eqnarray}
where $L_0$ is the typical luminosity for the mass $M_0$, $a$ and $b$ are model parameters for the peak luminosity, and the function $g(t_{\rm R})$ denotes the time evolution.
The peak of the luminosity is achieved at $t_{\rm R} = t_{\rm R}^{\rm peak}$, i.e., $g(t_{\rm R}^{\rm peak}) = 1$.
In the following, we distinguish the parameters for radio and neutrino emissions using the subscript, e.g., $(a_{\rm radio},b_{\rm radio})$ for radio, and $(a_{\nu},b_{\nu})$ for neutrino.
The radio luminosity, $L_{1.4}$, is measured at 1.4 GHz, while the neutrino luminosity, $L_{E_\nu}$, is measured at 100 TeV.
For a direct comparison with observations, $L_{\rm 1.4}$ is expressed in the unit of [W/Hz], while $L_{E_\nu}$ is in [GeV/s/GeV] (the number luminosity per unit energy multiplied by the energy).
The parameter set $(L_0,M_0)$ for the normalization is arbitrary.
Here, we adopt $(L_0,M_0) = (10^{24.5}~{\rm W/Hz},10^{14.9}~M_\odot)$ for radio \citep[e.g.,][]{Cassano2010}, and $(L_0,M_0) = (10^{40}~{\rm GeV/s/GeV},10^{15}~M_\odot)$ for neutrino emission.
\par

\begin{figure}
    \centering
    \plotone{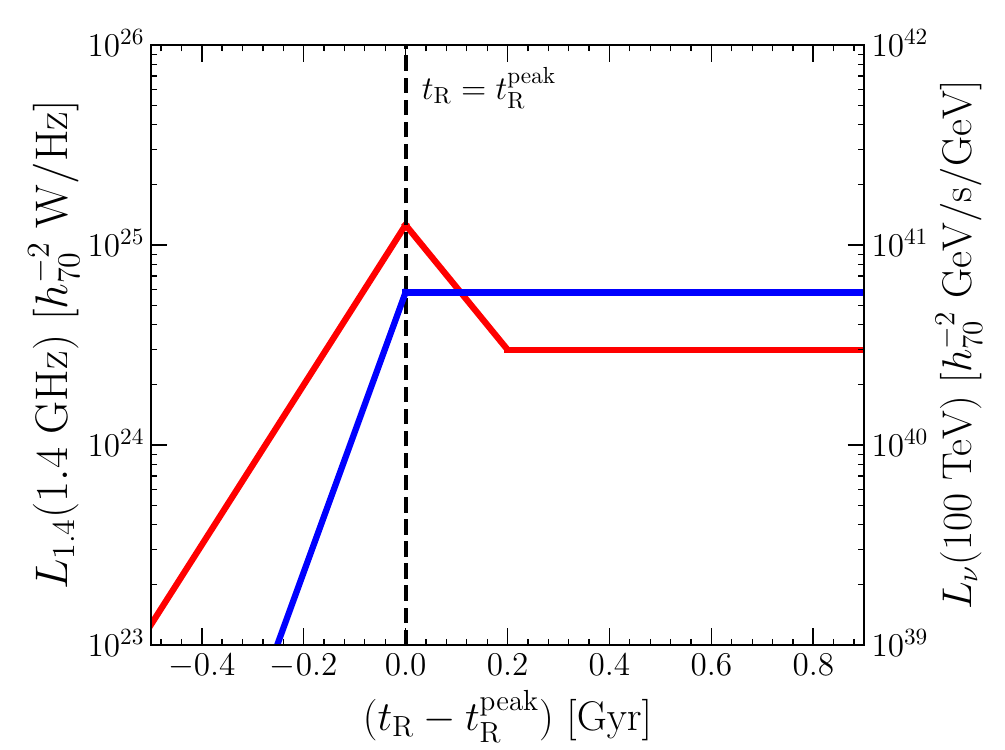}
    \caption{Time evolution of synchrotron luminosity at 1.4 GHz (red thick line, left axis) and neutrino luminosity at 100 TeV (blue thin line, right axis). 
    We consider a Coma-like cluster with the mass of $M_{500}=8.5\times10^{14}~M_\odot$ and adopt the re-acceleration timescale of $t_{\rm acc}=330~{\rm Myr}$.
    The spectral index of the primary CRPs is $\alpha_{\rm inj}=2.45$.
    Those trends are modeled with $g(t_{\rm R})$. The re-acceleration starts at $t_{\rm R}=0$ and ends at $t_{\rm R} = t_{\rm R}^{\rm peak}$ (dashed line).}
    \label{fig:L_time}
\end{figure}

The temporal evolution of the luminosities numerically obtained in our method is simplified as shown in Fig.~\ref{fig:L_time}, where we present an example case with $\alpha_{\rm inj}=2.45$ and $t_{\rm acc} \equiv p^2/4D_{pp} = 330~{\rm Myr}$, where $D_{pp}$ is the momentum diffusion coefficient in the FP equations.
Both synchrotron and neutrino luminosities increase with time during the re-acceleration, but the increasing rates are different.
In a $\sim\mu{\rm G}$ magnetic field, GHz emission is mainly produced by CREs with $\sim{\rm GeV}$ energies, which is produced through the $pp$ collision of $\sim10~{\rm GeV}$ CRPs.
The Coulomb cooling for CRPs suppresses the re-acceleration of GeV CRPs, while higher energy CRPs are efficiently re-accelerated without cooling.
Thus, the luminosity of 100~TeV neutrino evolves more rapidly than the GHz radio luminosity (see Fig.~2 in \citet{paperI}).
On the other hand, the gamma-ray luminosity at $\sim$ GeV evolves at a similar rate as the radio luminosity.
\par

%

As seen from Fig.~\ref{fig:L_time}, the temporal evolutions of the synchrotron and neutrino luminosities are different, so we adopt different functional forms of $g(t_{\rm R})$ in Eq.~(\ref{eq:Luminosity_model}) for each emission.
The temporal evolution of the radio luminosity at 1.4 GHz, $g_{1.4}(t_{\rm R})$, is given in \citet{paperII}.
For the neutrino emission, we adopt
\begin{eqnarray}
g_{\nu}(t_{\rm R}) =
\left\{
\begin{array}{cc}
    10^{3.5\left(\frac{t_{\rm R}-t_{\rm R}^{\rm peak}}{t_{\rm R}^{\rm peak}}\right)} & (t_{\rm R}-t_{\rm R}^{\rm peak}<0), \\
    1 & (0<t_{\rm R}-t_{\rm R}^{\rm peak}),  
\end{array}
\right.
\label{eq:h_nu}
\end{eqnarray}
with $t_{\rm R}^{\rm peak}=600~{\rm Myr}$ \citep[][]{paperII}.
\par


\subsection{Model Parameters}
\label{subsec:model_parameters}
\subsubsection{Fiducial Model}\label{subsec:fid.}
The parameters $(a_{\rm radio}, b_{\rm radio})$ and $\xi_{\rm RH}$ are constrained from the statistical properties of GRHs. 
In \citet{paperII}, we find the best fit values of $(a_{\rm radio}, b_{\rm radio})=(0.6, 3.5)$ and $\xi_{\rm RH} = 0.12$, with which the occurrence of the GRHs $f_{\rm RH}$ and its mass dependence 1.4 GHz are well reproduced \citep[e.g.,][]{2021A&A...647A..50C}.
We call the model with these parameters the ``fiducial model".
\par


From the simulations in \citet{paperII}, we find $b_{\rm radio}\sim 3$ for $D_{pp}\propto M_{500}^{1/3}$.
This mass dependence in the diffusion coefficient is expected in the re-acceleration by the transit-time damping (TTD) with the compressible turbulence \citep[e.g.,][]{BL07}.
With this mass dependence, the parameters for the neutrino LM relation are found to be $(a_\nu,b_\nu)=(0.93, 5.1)$.
The difference between $b_{\rm radio}$ and $b_\nu$ reflects the difference in the spectral evolution of $\sim10$GeV CRPs and $\sim1$PeV CRPs as mentioned in Section~\ref{subsec:L_evolution}.
\par


\begin{table}[h]
    \centering
    \caption{Parameters for the fiducial model (fid.) and the optimistic model (opt.). The parameters $\xi_{\rm RH}$ and $(a_{\rm radio}, b_{\rm radio})$ are constrained by the radio observation. Given $b_{\rm radio}$, the relevant mass dependence of $D_{pp}$ and the neutrino parameters $(a_\nu, b_\nu)$ are shown.}
    \begin{tabular}{ccccc}
    \hline
      model & $D_{pp}$ & $\xi_{\rm RH}$ & $(a_{\rm radio}, b_{\rm radio})$ &  $(a_\nu, b_\nu)$\\
      \hline
       fid. &$\propto M^{1/3}$& 0.12 & (0.60, 3.5)   & (0.93, 5.1) \\
       opt. & $\propto M^{0}$  & 0.30 & (0.60, 1.5) &   (0.73, 1.7)  \\
    \hline
    \end{tabular}
    \label{tab:params}
\end{table}

\subsubsection{Optimistic Model}\label{subsec:opt.}
Since the LM relations are normalized at $M_{500}\approx 10^{15}M_\odot$, which is the typical mass of observed GRHs,
the contribution to the background radiation from smaller mass clusters ($M_{500}\simeq10^{14}M_\odot$) increases when a shallower LM relation is adopted.
To discuss the upper limits for the contributions to the ARCADE-2 excess and IceCube neutrino, we test the ``optimistic model" with smaller $b_{\rm radio}$ and $b_{\nu}$.
\par

In the optimistic model, we assume $(a_{\rm radio},b_{\rm radio})=(0.6,1.5)$.
In this case, the reproductivity of the statistical properties for GRHs becomes worse than the fiducial model.
For example, the threshold mass ratio $\xi_{\rm RH}$ needs to be larger than the fiducial model to fit the observed $f_{\rm RH}$ in the high-mass range ($8\times10^{14}M_\odot<M_{500}<1.2\times10^{15}M_\odot$). 
In the low-mass range ($5\times10^{14}M_\odot<M_{500}<8\times10^{14}M_\odot$), it becomes $f_{\rm RH} = 0.52\pm0.06$, which is larger than the observed value ($f_{\rm RH} = 0.37\pm 0.02$) \citep[e.g.,][]{2021A&A...647A..50C}.
Thus, we consider that $b_{\rm radio}<1.5$ is excluded due to the mass trend of the $f_{\rm RH}$.
In addition, such a small value of $b_{\rm radio}$ is inconsistent with the LM relation observed at 150 MHz \citep[][]{Cuciti_2023}.
\par

Solving the FP equation, we study the mass scaling of $D_{pp}$ that corresponds to $b_{\rm radio}\sim1.5$.
We find a similar scaling, $b_{\rm radio} = 1.8$, in the model with $D_{pp}\propto M^0$.
In this case, we obtain the coefficients for the neutrino LM relation as $(a_\nu,b_\nu) = (0.73,1.67)$.
Thus, we adopt the parameters as summarized in Table \ref{tab:params} for the optimistic model.
\par

In the ``accretion shock" models \citep[e.g.,][]{Keshet:2002sw,Murase2008,Zandanel2015}, the relation $L_{E_\nu}\propto M^{5/3}$, which corresponds to the mass-independent neutrino production efficiency for inelastic $pp$ collisions, is often adopted.
Our optimistic model corresponds to this assumption, because we assumed $Q_p\propto M^{2/3}$ for the CR injection rate and $D_{pp}\propto M^0$, and neglected the effects of spatial diffusion \citep[see][for the details]{paperII}.
\par

\subsection{Luminosity Functions}
\label{subsec:LF}

\begin{figure*}
    \centering
    \plottwo{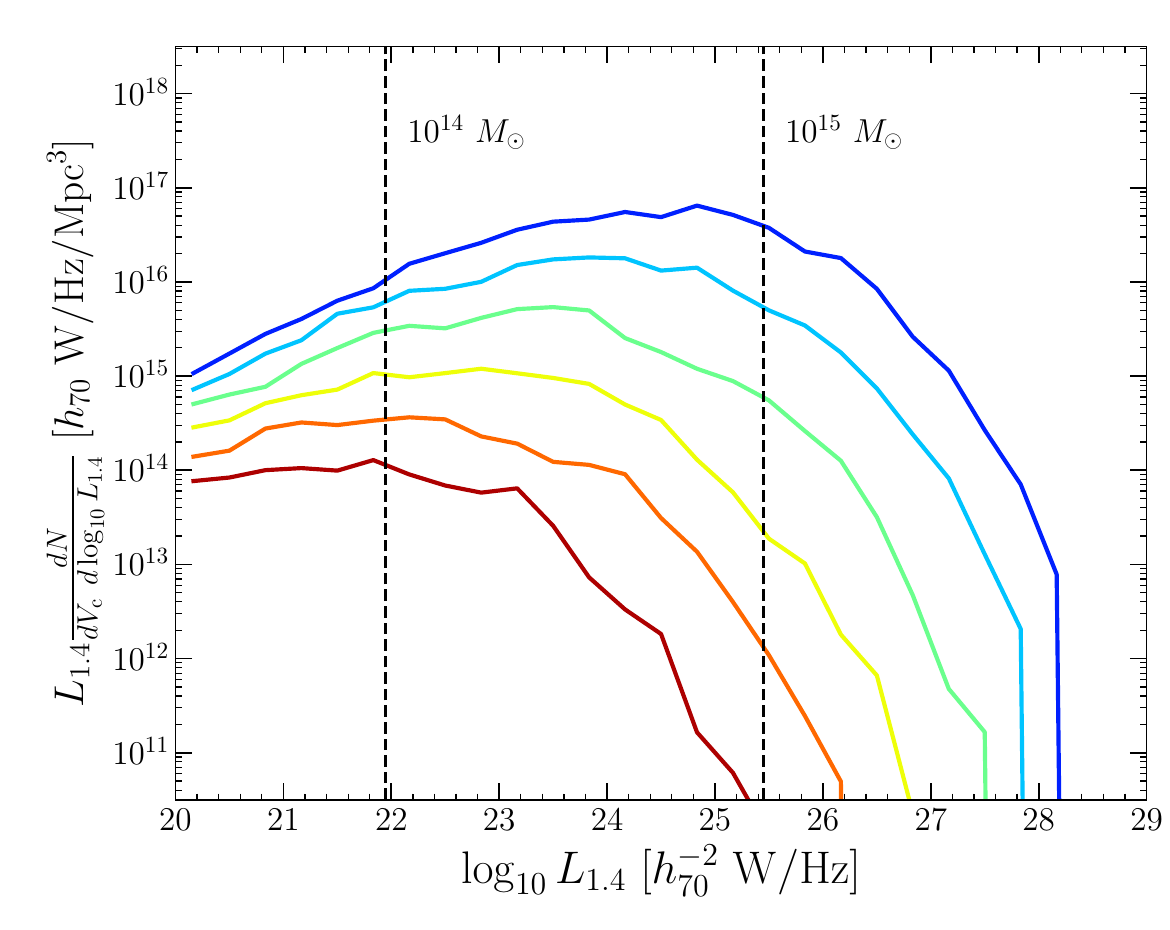}{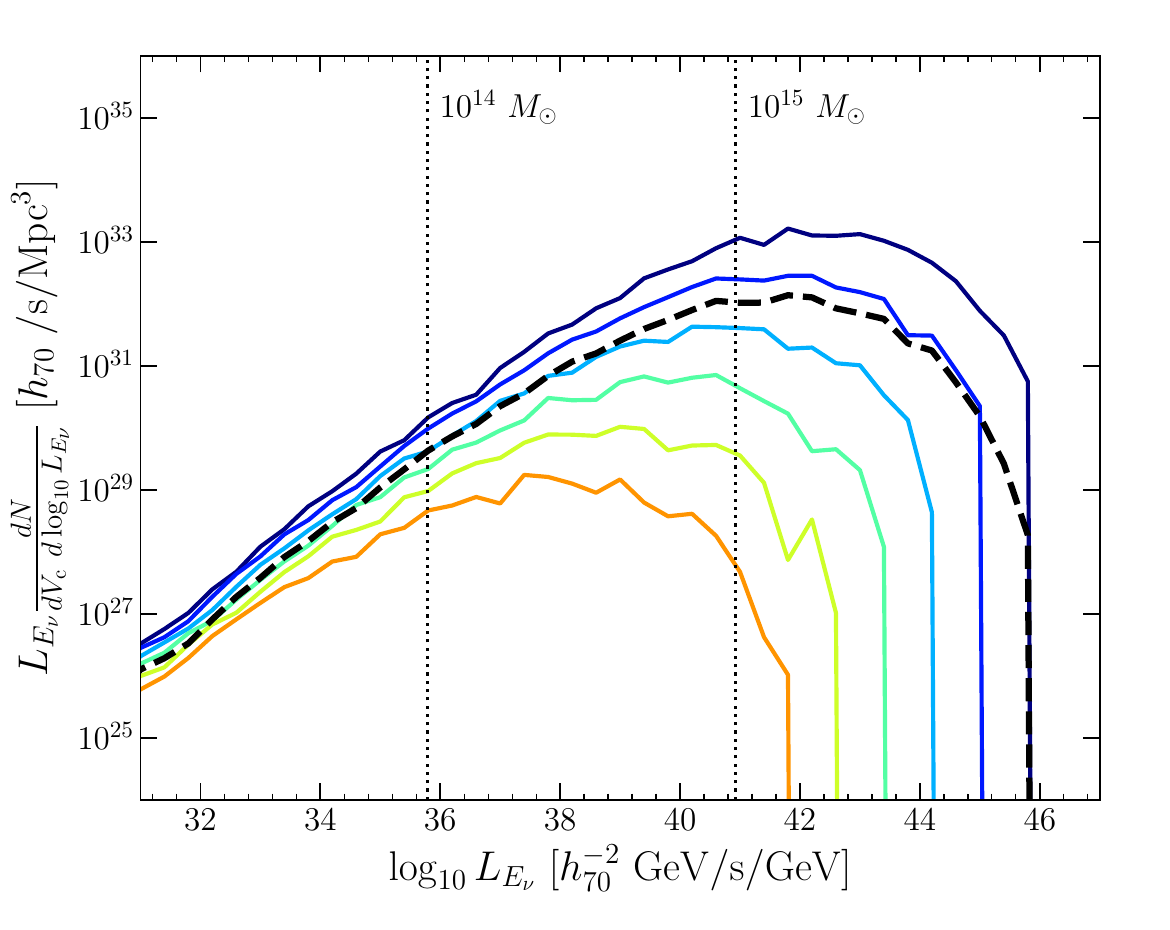}
    \caption{Luminosity functions for radio (1.4 GHz) and neutrino (100 TeV) emissions in the fiducial model. The LFs are multiplied by the luminosity. The redshift evolution from $z =1$ to $z=0$ (every 0.2, from red to blue) are shown. The vertical dashed lines show the luminosities corresponding to the mass $M_{500} = 10^{13}M_\odot,10^{14}M_\odot$, and $10^{15}M_\odot$ calculated with the $L-M$ relation at $z=0$. The black dashed line in the right panel shows the LF averaged within $z<z_{\rm max}$, where the sources produce 99\% of the total neutrino background of the model.}
    \label{fig:LF}
\end{figure*}

We obtain the LFs for radio and neutrino emissions by counting the number of GCs in the merger tree with a certain luminosity at a given redshift.
In Fig.~\ref{fig:LF}, we show the LFs for the synchrotron (1.4~GHz) and neutrino (100~TeV) emissions in our fiducial model. 
The LFs are multiplied by the luminosity ($L_{1.4}$ or $L_{E_\nu}$) in Fig.~\ref{fig:LF} so that one can see the relative contribution to the isotropic background from each luminosity bin.
The dashed line shows the neutrino LF averaged within $z<z_{\rm max}$, where $z_{\rm max}$ is the maximum redshift within which 99\% of the neutrino background intensity is produced (Sect.~\ref{sec:contribution_backgound}).
In the fiducial model, we find $z_{\rm max}\approx 0.4$.
We define the effective luminosity, $L^{\rm eff}$, as the luminosity that maximize $L(dN/dV_{\rm c}d\log L)$ \citep[e.g.,][]{Murase_Waxman2016}.
The averaged LF implies that the effective luminosity at 100 TeV becomes $L_{E_\nu}^{\rm eff} = 10^{42}~{\rm GeV/s/GeV}$ ($E_{\nu_\mu} L_{E_{\nu_\mu}}^{\rm eff} \approx 5.3\times10^{43}~{\rm erg/s}$ for muon neutrinos), which corresponds to the mass $M_{500}\approx1.5\times10^{15}M_\odot$.
The effective source number density is calculated as 
\begin{equation}\label{eq:n_eff}
    n^{\rm eff} = \frac{1}{L_{E_\nu}^{\rm eff}}\int (d \log_{10} L_{E_\nu}) L_{E_\nu} \frac{dN}{dV_{\rm c}d\log_{10}L_{E_\nu}},
\end{equation}
which is estimated to be $n^{\rm eff} \approx 4 \times10^{-9}~{\rm Mpc}^{-3}$ in the fiducial model, implying that the contribution is dominated by a few nearby galaxy clusters.
\par

In the optimistic model, the effective luminosity becomes smaller due to the shallower LM relations.
We find $L_{E_\nu}^{\rm eff} = 2.7\times10^{38}~{\rm GeV/s/GeV}$ ($E_\nu L_{E_\nu}^{\rm eff}\approx1.4\times10^{40}$ erg/s for muon neutrino) and $z_{\rm max} \approx 1.1$.
That luminosity corresponds to the mass of $M_{500} \approx 1.4\times10^{14}M_\odot$.
The effective source number density for the neutrino background becomes $n^{\rm eff}_{\nu}=1.0\times10^{-5}~{\rm Mpc}^{-3}$.
The number of the GRH observation around this mass scale \citep[e.g.,][]{2021ApJ...914L..29B} is not significant due to the limited sensitivity of current instruments.

Considering the non-detection of multiple muon neutrino events from a single source in the IceCube observation, one can put an upper limit on the source number density $n^{\rm eff}_\nu$ (Eq.(~\ref{eq:n_eff})) for a given neutrino luminosity \citep[e.g.,][]{Murase_Waxman2016}.
According to \citet{Murase_Waxman2016}, the upper limits for neutrino sources with $E_{\nu_\mu} L_{E_{\nu_\mu}}^{\rm eff} \approx 5.3\times10^{43}~{\rm erg/s}$ and $E_\nu L_{E_\nu}^{\rm eff}\approx1.4\times10^{40}$ erg/s and are $n^{\rm eff}_{\nu}<1.6\times10^{-9}~{\rm Mpc}^{-3}$ and $n^{\rm eff}_{\nu}<2\times10^{-4}~{\rm Mpc}^{-3}$, respectively.
The effective densities implied from our fiducial and optimistic models fall below those limits, which is consistent with the no significant detection from nearby clusters with the current IceCube sensitivity \citep[][]{IceCube_2013_GC,IceCube_2022_stack}.

\section{Contribution to All-Sky Fluxes}
\label{sec:contribution_backgound}
In this section, we discuss the contribution of massive GCs to the isotropic background fluxes of radio waves and high-energy neutrinos. In the previous section, we obtained the LFs at a given frequency (or energy). 
For the spectrum, we assume that all the clusters share the same spectral shape as that in the Coma cluster shown in Fig.7 of \citet[][]{paperI}.
The neutrino spectrum can be approximated as a single power-law, while the synchrotron spectrum shows a steepening around 1.4 GHz.
We do not consider any redshift or mass dependence in the spectral shape.
Note that the spectral index of the Coma GRH below 1.4 GHz, $\alpha_{\rm syn}=-1.22$, is typical among the observed population \citep[e.g.,][]{Brunetti_Jones_review,review_vanWeeren}.
\par

The mean intensity of the all-sky isotropic background flux ($dE/d\nu dtdAd\Omega$) can be calculated by
\begin{eqnarray}\label{eq:isotropic_intensity}
E_\nu\Phi_\nu & = &\frac{1}{4\pi}\int dz \frac{dV_{\rm c}}{dz}\int dL_{\rm \nu_{\rm i}} \frac{dN}{dL_{\nu_{\rm i}}dV_{\rm c}}L_{\nu_{\rm i}}\frac{(1+z)}{4\pi D_{\rm L}^2(z)}e^{-\tau(\nu,z)}, \nonumber \\
& = & \frac{1}{4\pi}\int dz \frac{c}{(1+z)H(z)}\int dL_{\nu_{\rm i}}\frac{dN}{dL_{\nu_{\rm i}}dV_{\rm c}}L_{\nu_{\rm i}}e^{-\tau(\nu,z)},
\end{eqnarray}
where $V_{\rm c}(z)$ is the comoving volume at redshift $z$, $\nu_{\rm i} = (1+z)\nu$ is the frequency at the source, $\frac{dN}{dL_{\nu_{\rm i}}dV_{\rm c}}$ is the luminosity function, $D_{\rm L}(z)$ is the luminosity distance, and $\tau(\nu,z)$ is the optical depth.
We assume that GCs are distributed isotropically and do not consider the anisotropy in the intensity.
We adopt $\tau(\nu,z)=0$ for the radio and neutrino backgrounds. 
For the neutrino background, we calculate the intensity per unit energy ($dE/dEdtdAd\Omega$) rather than that per unit frequency.
\par


\subsection{ARCADE-2 excess radio emission}
\label{subsec:ARCADE}
\begin{figure}
    \centering
    \plotone{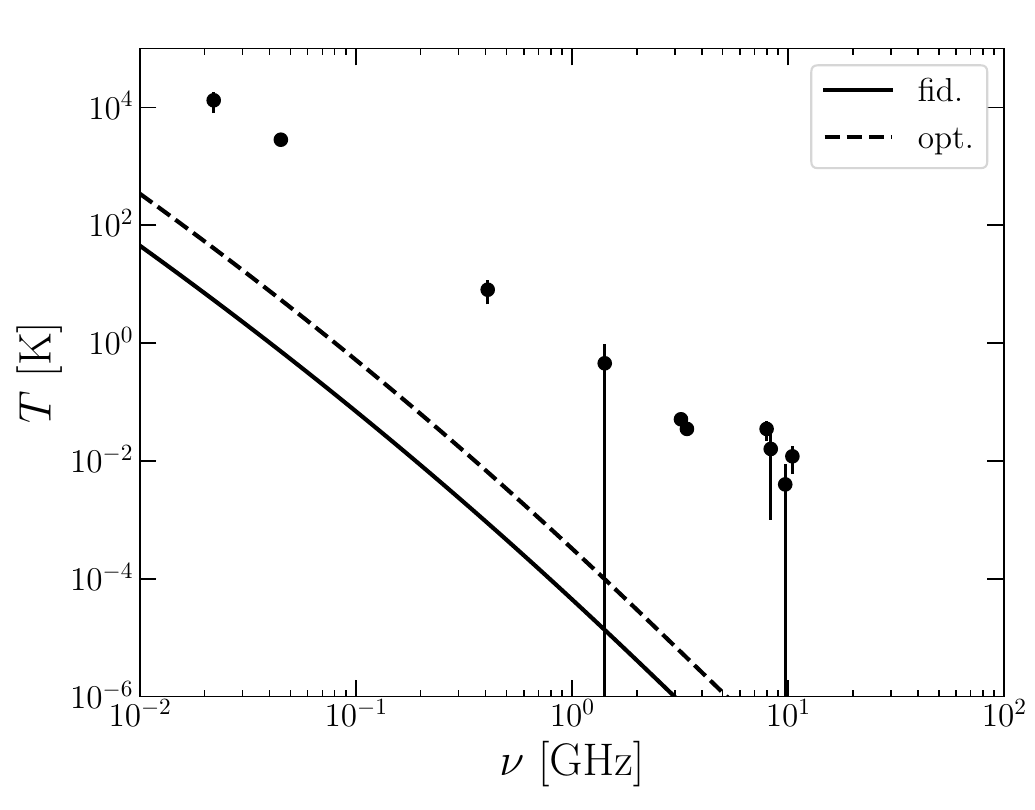}
    \caption{Brightness temperature of the isotropic radio background expected from galaxy clusters. The solid and dashed lines show the results of the fiducial and the optimistic models, respectively. The data points show the measured radio background after subtracting off the CMB temperature \citep[][]{ARCADE2_Fixsen_2011ApJ...734....5F}.}
    \label{fig:ARCADE2}
\end{figure}

Several candidate sources for the ARCADE-2 excess have been examined so far, such as AGNs \citep[][]{Draper_2011}, supernovae of massive pop III stars \citep[][]{Biermann_2014,Jana_2019}, and the Galactic halo \citep[][]{Subrahmanyan_Cowsik_2013}.
Importantly, \citet{Vernstrom_2011} showed that the excess cannot be explained by the sum of known extragalactic point sources, and that a large number of new point sources with 1 GHz fluxes smaller than $\sim10~\mu{\rm Jy}$ is required.
\citet{Fang_Linden:2016dga} claimed that the diffuse radio emission in GCs could explain a large part of the radio excess. Their model is not constrained by the high isotropy of the signal measured at small angular scales \citep[][]{Holder_2014ApJ...780..112H}.

\par

We calculate the radio background from GCs, using Eq.~(\ref{eq:isotropic_intensity}) and the LF shown in Fig.~\ref{fig:LF}.
In Fig.~\ref{fig:ARCADE2}, we compare the spectra of the radio background from GCs with the ARCADE-2 excess.
The data points are taken from \citet{ARCADE2_Fixsen_2011ApJ...734....5F}, which includes the result of low-frequency surveys below 3 GHz \citep[][]{Roger_1999A&AS..137....7R,Maeda_1999A&AS..140..145M,Haslam_1982A&AS...47....1H,Reich_1986A&AS...63..205R}.

We find that the contribution from GCs is not significant for all frequencies between $0.01~{\rm GHz}<\nu<10~{\rm GHz}$.
This result is in contrast to the previous estimate by \citet[][]{Fang_Linden:2016dga}, in which
almost 100\% of ARCADE-2 excess is explained by the emission from GCs with the re-acceleration model of \citet{Fujita_2003ApJ...584..190F}.
While \citet{Fujita_2003ApJ...584..190F} succeeded in reproducing the typical value of the bolometric synchrotron luminosity found in $M\sim10^{15}M_\odot$ clusters ($L_{\rm syn}\sim 3\times10^{40}$ erg/s)\citep[e.g.,][]{Cassano_2013ApJ...777..141C}, \citet{Fang_Linden:2016dga} used a larger value $L_{\rm syn}\sim 3\times10^{42}$ erg/s for the same mass.
This discrepancy might be originated from the different assumption on the turbulent velocity, i.e., the parameter $\eta_v$ in \citet[][]{Fang_Linden:2016dga}. 
\par

On the other hand, our model is based on the theory of particle acceleration through TTD, whose consistency with radio observations are extensively discussed in previous studies \citep[e.g.,][]{BL07,Pinzke2017,paperII}.
Based on this model, we conclude that massive galaxy clusters are not the dominant source of the ARCADE-2 excess.
\par

We obtain a radio LF similar to Fig.~\ref{fig:LF} in the ``primary scenario", where the seed population is dominated by the primary CREs \citep[][]{paperII}. 
The conclusion in this section does not depend on the seed origin, i.e., primaries or secondaries, as long as the radio LF at 1.4 GHz is consistent with the observation.



\subsection{IceCube neutrino emission}
\label{subsec:IceCube}
\begin{figure}
    \centering
    \plotone{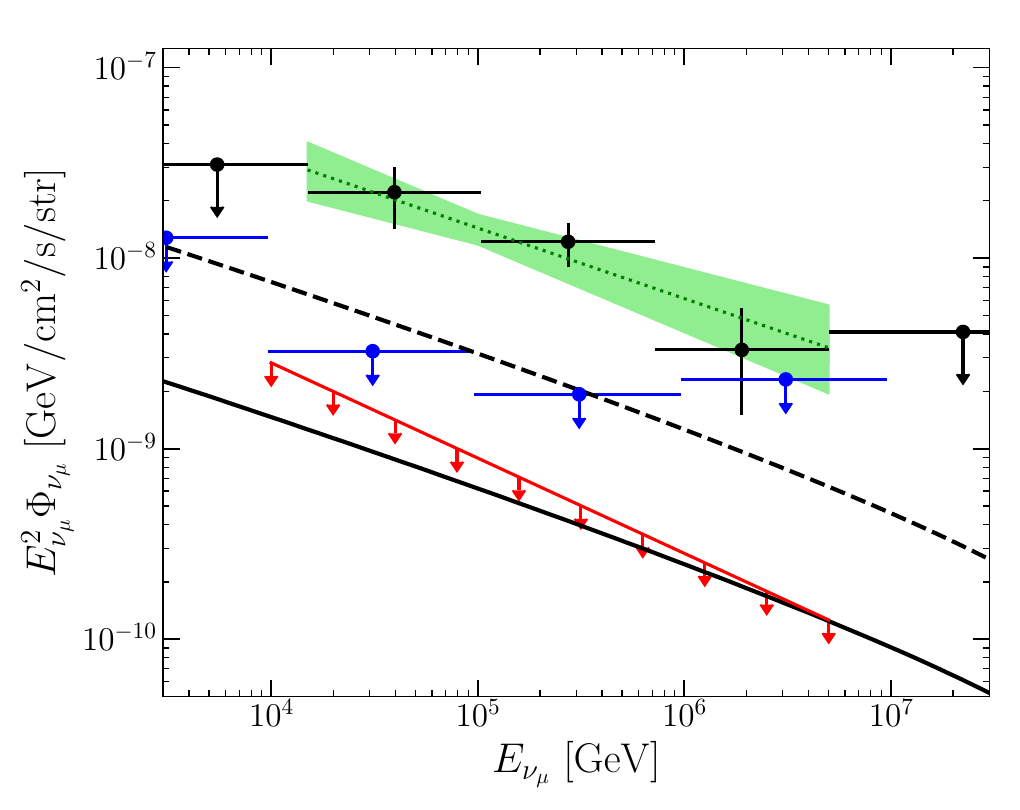}
    \caption{Spectra of the diffuse high-energy muon neutrino flux. The data points and green band show the all-sky muon neutrino intensity estimated with IceCube track events \citep[][]{IceCube_2022_muontrack}.
    The red line shows the 90\% CL upper limit derived from the stacking analysis of {\it Planck}-SZ clusters using the distance-weighting scheme ($L_{E_\nu}\propto M^0$) \citep[][]{IceCube_2022_stack}.
    The blue arrows show the upper limits in quasi-differential energy bins for the distance-weighting scheme \citep[][]{IceCube_2022_stack}.
    The solid and dashed lines show the results of the fiducial and the optimistic models, respectively.}
    \label{fig:IceCube}
\end{figure}

In Fig.~\ref{fig:IceCube}, we show the contributions to the isotropic neutrino background flux.
We derive the per-flavor flux by multiplying the all-flavor flux by 1/3, assuming the flavor ratio of $\nu_e:\nu_\mu:\nu_\tau  = 1:1:1$.
The expected flux is $\approx5\%$ of the observed intensity around 100 TeV in the fiducial model, while it reaches $\approx18\%$ in the optimistic model.
The result in the fiducial model is comparable to our previous estimate \citep[][]{paperI}, where we only considered the contribution from Coma-like clusters ($M_{500}\approx8.5\times10^{14}M_\odot$).
\par
The red line shows the 90\% confidence level (CL) upper limit for an $E_\nu^{-2.5}$ spectrum given by the stacking analysis of {\it Planck}-SZ clusters using 9.5 years of muon-neutrino track events \citep[][]{IceCube_2022_stack}.
To assess the completeness of the {\it Planck} catalog, the ``distance-weighting" scheme is adopted, where the neutrino flux from each cluster is proportional to the inverse of distance squared and independent of the mass ($F_{\nu}\propto M^0/D_{\rm L}^2$) \citep[see][for details]{IceCube_2022_stack}.
The flux in the fiducial model is comparable to this limit, which means that the stacking analysis of the neutrino events is sensitive enough to test the secondary dominant model for the GRHs (see Sect.~\ref{subsec:future_obs}).
\par

Note that the upper limit depends on the weighting scheme assumed in the analysis. For example, the limit from the mass-weighting ($L_{E_\nu} \propto M^1$) scheme corresponds to the fractional contribution of 4.6\% to the total flux at 100 TeV, while the limit from the distance-weighting (Fig.~\ref{fig:IceCube}) corresponds to 6.2\% fractional contribution \citep[][]{IceCube_2022_stack}.
If we adopt larger weights on nearby massive clusters, which are almost completely covered by the catalog, the upper limit could be slightly more stringent than the estimates above.
The red line in Fig.~\ref{fig:IceCube} should be regarded as a conservative upper limit for our fiducial and optimistic models.
\par

The diffuse flux in the optimistic model exceeds the limit derived with the distance weighting.
A similar level of flux was predicted in the so-called ``accretion-shock" model \citep[][]{Murase2008,Fang_Olinto2016,Hussain:2021dqp}, where the same scale for the neutrino LM relation ($L_{E_\nu} \propto M^{5/3}$) is assumed.
The blue arrows in Fig.~\ref{fig:IceCube} show the upper limit in quasi-differential energy bins for the distance weighting, which is derived by assuming an $E_\nu^{-2}$ spectrum across each one-decade energy bin.
As discussed in \citet{IceCube_2022_stack}, this limit is not deep enough to exclude those models.
Thus, the optimistic model ($L_{E_\nu} \propto M^{5/3}$) is in tension with the stacking limit for an $E_\nu^{-2.5}$ spectrum, but it is not completely excluded due to the uncertainty in the spectral index.

Note that we claim that the radio luminosity should follow $L_{\rm 1.4}\propto M^{2}$ in the optimistic scenario, and such a model is in tension with the current statistics of GRHs (Section~\ref{subsec:model_parameters}).
\par

\section{Multi-messenger constraints on the electron-to-proton ratio and magnetic field}\label{sec:constraint}

As discussed in Sect.~\ref{subsec:IceCube}, the neutrino upper limit is deep enough to constrain our secondary-dominant model for GRHs.
In this section, we discuss the possible constraints on the model parameters in the re-acceleration model from the neutrino upper limit, and compare them with the limits from the gamma-ray upper limit discussed in previous studies.
\par

The statistical properties of GRHs can be well reproduced with the fiducial ($D_{pp}\propto M^{1/3}$) model, where the LFs have peaks at the luminosities corresponding to $M_{500}\approx 10^{15}M_\odot$ and those clusters most effectively contribute to the neutrino background.
Since our models are normalized with the GRH observations, we focus on the ratio of the neutrino luminosity to the radio luminosity $L_{E_\nu}/L_{\rm radio}$ at $10^{15}M_\odot$.
As in the previous sections, we consider the radio luminosity at 1.4 GHz and the neutrino luminosity at 100 TeV.
\par

In our re-acceleration model, the radio and neutrino fluxes depend on parameters such as acceleration time $t_{\rm acc}$, duration of the re-acceleration $\Delta T$, spectral index $\alpha_{\rm inj}$, magnetic field $B$, and electron-to-proton ratio of primary CRs $f_{\rm ep}$ \citep[][]{paperI}.
In this section, we constrain $f_{\rm ep}$ and $B$, fixing $\alpha_{\rm inj} = 2.45$ and $\Delta T/t_{\rm acc} = 2.8$ \citep[][]{paperII}. 
We assume the magnetic field profile as $B \propto n_{\rm ICM}^{\eta_B}$.
As in the previous sections, the ICM density profile of the Coma cluster is adopted.
We also assume that $t_{\rm acc}$ is short enough to balance the cooling rate of CREs at the energy corresponding to the synchrotron frequency of 1.4 GHz for a given $B$. 

\par

The ratio $L_{E_\nu}/L_{\rm radio}$ depends on the magnetic field, which is one of the uncertain parameters.
Approximating the CRE spectrum as a single power-law, the synchrotron luminosity at a given frequency (e.g., 1.4 GHz) scales as $L_{\rm radio}\propto B^{-(q+1)/2}$, where $q$ is the spectral index of CREs \citep[][]{Rybicki:847173}.
The estimate of $q$ is not straightforward, because the spectral shape after re-acceleration generally deviates from the single power-law and depends on $B$.
To simplify our estimate method, we consider the case where $q$ can be approximated as $q \approx 3.45$ as expected from the cooling of CREs injected with $\alpha_{\rm inj} = 2.45$.
This spectral index is roughly consistent with the radio spectral index typically observed in GRHs.
\par


The CRE injection rate is written as $Q_{\rm e}(p) = Q_{\rm e}^{\rm pri}(p)+Q_{\rm e}^{\rm sec}(p)$, where $Q_{\rm e}^{\rm pri}(p)$ and $Q_{\rm e}^{\rm sec}(p)$ are the injection rates of primary and secondary CREs, respectively. The number of seed CREs in the equilibrium state before the re-acceleration can be calculated as $N_{\rm seed} \approx \int Q_{\rm e}(p)dp/ |dp/dt|_{\rm cool}$, where $|dp/dt|_{\rm cool}$ is the cooling rate of CREs.
For the synchrotron and inverse-Compton (IC) cooling, $|dp/dt|_{\rm cool}\propto p^2(B^2+B_{\rm cmb}(z)^2)$, where $B_{\rm cmb} = 3.2(1+z)^4~\mu{\rm G}$ is the equivalent magnetic field for the IC cooling with CMB photons. For a fixed $Q_{\rm e}(p)$, we obtain the magnetic field dependence of the radio luminosity as \citep[e.g.,][]{Brunetti_Jones_review}

\begin{equation} \label{eq:Lsyn_B}
    L_{\rm radio}\propto \frac{B^{\frac{q+1}{2}}}{(B^2+B^2_{\rm cmb}(z))}.
\end{equation}
\par

Assuming the same spectral indexes $\alpha_{\rm inj}$ for the primary CREs and CRPs, the spectral index of secondary CREs is very close to $\alpha_{\rm inj}$.
In this case, the ratio between $Q_{\rm e}^{\rm pri}$ and $Q_{\rm e}^{\rm sec}$ is directly related to the electron-to-proton ratio $f_{\rm ep}$ as
\begin{equation} \label{eq:pri_sec}
    \frac{Q_{\rm e}^{\rm sec}}{Q_{\rm e}^{\rm pri}+Q_{\rm e}^{\rm sec}} \propto  \frac{f_{\rm ep}^0}{f_{\rm ep}+f_{\rm ep}^0},
\end{equation}
where $f_{\rm ep}=f_{\rm ep}^0 \approx 10^{-2}$ is the value at which the secondary CREs contribute comparably to the primary CREs.
Since the neutrino luminosity is proportional to $Q_{\rm e}^{\rm sec}$, the dependence of $L_{E_{\nu}}/L_{\rm radio}$ on $f_{\rm ep}$ can be expressed with with Eq.~(\ref{eq:pri_sec}).
Note that the neutrino flux is not affected by the change of $B$, since CRPs are free from the radiative cooling.
\par

\par

\begin{figure}
    \centering
    \plotone{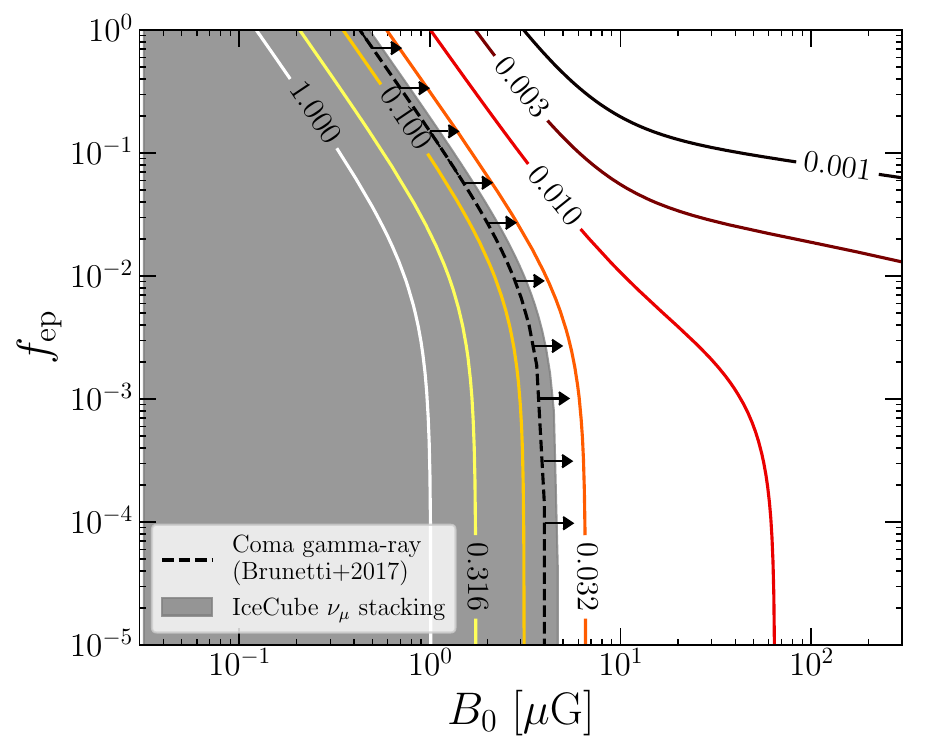}
    \caption{Constraints on $f_{\rm ep}$ and the central magnetic field $B_0$. The shaded region is excluded from the neutrino upper limit for $E^{-2.5}$ spectrum shown in Fig.~\ref{fig:IceCube}. 
    We have assumed that the neutrino background is dominated by the massive clusters with $M_{500}\approx 10^{15} M_\odot$.
    The contour shows the fractional contribution to the all-sky neutrino intensity calculated with Eq.~(\ref{eq:Lnu_Lradio}) with $q = 3.45$ and $z = 0.2$.
    We fix $\eta_{B} = 0.5$ for the magnetic field profile, and the ICM density profile of the Coma cluster is adopted.
    The dashed line with arrows shows the minimum value of $B_0$ for the Coma cluster constrained by the Fermi-LAT upper limit in the re-acceleration model \citep[][]{Brunetti2017}. 
    Note that the parameters in \citet{Brunetti2017} is similar to our model: $\eta_B = 0.5$, $f_{\rm ep} = 0$, $t_{\rm acc} = 260$ Myr, and $\Delta T/t_{\rm acc} \approx 2.8$.
    The limit is scaled with $f_{\rm ep}$ as shown in Eq.~(\ref{eq:pri_sec}).}
    \label{fig:fep_B}
\end{figure}

From Eqs.~(\ref{eq:Lsyn_B}) and (\ref{eq:pri_sec}), $L_{E_\nu}/L_{\rm radio}$ depends on $f_{\rm ep}$ and $B$ as
\begin{equation}\label{eq:Lnu_Lradio}
    \frac{L_{E_\nu}}{L_{\rm radio}} \propto \left(\frac{f_{\rm ep}^0}{f_{\rm ep} + f_{\rm ep}^0}\right){B^{-\frac{q+1}{2}}\left(1+\left(\frac{B}{B_{\rm cmb}(z)}\right)^2\right)}.
\end{equation}

Fig.~\ref{fig:fep_B} shows the constraints on $f_{\rm ep}$ and $B$ from the neutrino upper limit.
The contour is calculated with Eq.~(\ref{eq:Lnu_Lradio}), which is normalized such that the fractional contribution to the total neutrino intensity becomes 5\% for $B_0 = 4.7~\mu{\rm G}$ and $f_{\rm ep} = 0$, i.e., the fiducial model in Sect.~\ref{subsec:IceCube}.
The shaded region is excluded from the neutrino upper limit with the distance-weighting scheme, which corresponds to the contribution of 6.2\% (Sect.~\ref{subsec:IceCube}).
We find that the secondary-dominant model ($f_{\rm ep}\ll f_{\rm ep}^0$) with $B<1~\mu{\rm G}$ is tension with the neutrino observation, but it is allowed for large magnetic fields.

Another important constraint on the CRPs in the ICM is available from the non-detection of gamma-rays from nearby clusters \citep[e.g.,][]{Jeltema_2011,Ackermann_2014,Zandanel2015,Ackermann2016_Coma,Brunetti2017}. 
The deep gamma-ray limit on the Coma cluster with {\it Fermi}-LAT is in tension with the pure hadronic origin of the RH without the re-acceleration, assuming the magnetic field implied from the RM observation \citep[][]{Bonafede2010}.
Although the tension is alleviated in the presence of the re-acceleration, \citet{Brunetti2017} found that the magnetic field should be larger than $B_0\gtrsim 4\mu{\rm G}$ when $\alpha_{\rm inj}= 2.45$, $f_{\rm ep} = 0$ (only secondaries) and $\eta_B = 0.5$.
This limit can be directly compared with the neutrino limit in Fig.~\ref{fig:fep_B}, since the parameters for the re-acceleration in \citet{Brunetti2017}, $t_{\rm acc} = 260$ Myr and $\Delta T/t_{\rm acc} = 2.77$, is almost the same as ours (see Sect.~\ref{sec:LF}).
The dashed line in Fig.~\ref{fig:fep_B} is drawn by extrapolating the limit by \citet{Brunetti2017} at $f_{\rm ep}=0$ using Eq.~(\ref{eq:pri_sec}).

\par

We find that the stacking analysis of the IceCube neutrino provides a limit comparable to the limit solely from the Coma cluster.
The central magnetic field measured with RM ranges $B_0 \approx 1- 10\mu{\rm G}$ \citep[][]{Govoni_2017}, and the upper limits still allow the secondary scenario within this range.
The limits become less stringent if one assumes a flatter magnetic field profile ($\eta_B<0.5$), as demonstrated in \citet{Brunetti2017}.
A neutrino limit as deep as 1\% of the IceCube level would significantly constrain the CRP content in the ICM and give a crucial hint to the origin of the seed CREs for the re-acceleration.
\par

\section{Future Observational Tests}\label{subsec:future_obs}

\begin{figure}
    \centering
    \plotone{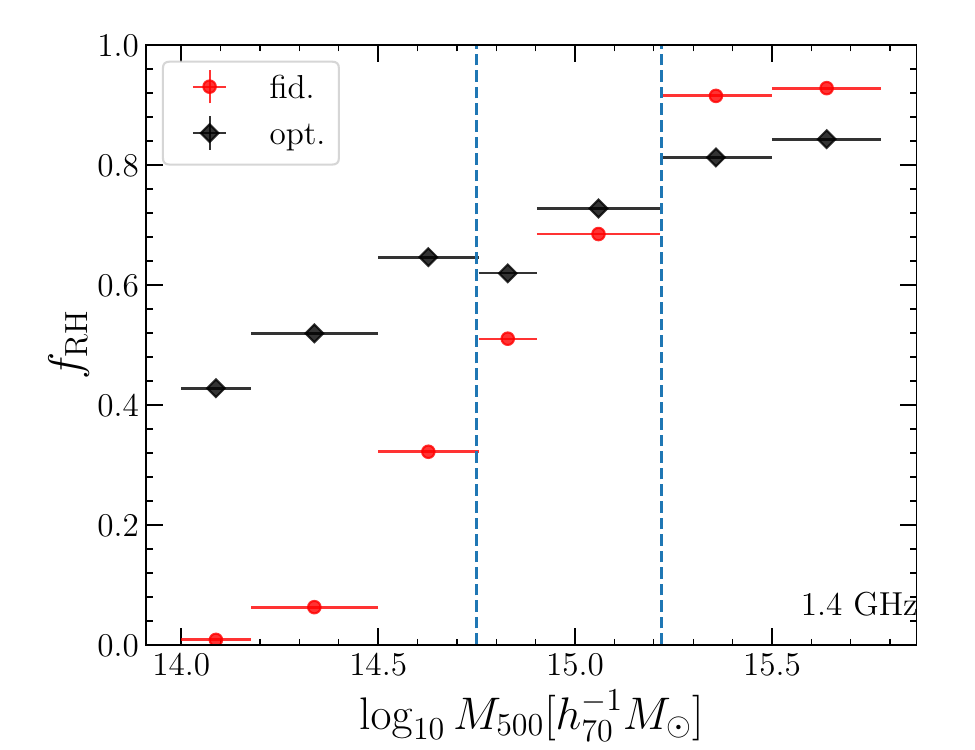}
    \caption{Fraction of radio halos detectable with the ASKAP sensitivity at 1.4 GHz. The black and red points show the fraction predicted in the fiducial and optimistic models, respectively. The dashed line shows the mass range studied in previous observations \citep[][]{2021A&A...647A..50C}.
    }
    \label{fig:fRH_ASKAP}
\end{figure}

Both our fiducial and optimistic models are compatible with current observations of GRHs. Indeed, both can fit the observed increasing trend of $f_{\rm RH}$ with cluster mass \citep[][]{Cuciti_2015A&A...580A..97C,2021A&A...647A..50C}.
However, the difference between those two models becomes apparent when comparing the RH fraction in the lower mass range, which can be tested with future observations with higher sensitivity.
Importantly, the upper limit from the stacking analysis on neutrino emission from galaxy clusters is deep enough to constrain our models \ref{subsec:IceCube}.
In this section, we discuss the possible constraints on the re-acceleration model of GRHs by the combination of stacking analysis and future radio halo surveys.
For simplicity, we consider the neutrino upper limit of $\sim5\%$ to the diffuse flux derived under the physically motivated weighting schemes (blue line in Fig.~\ref{fig:IceCube}).
\par

The future radio survey with the Australian SKA Pathfinder (ASKAP) will improve the statistics, especially at the lower mass range.
The fiducial scenario predicts that $f_{\rm RH}$ observed with the ASKAP sensitivity would be $f_{\rm RH}\approx0.05$ in $10^{14}M_\odot<M_{500}<10^{14.5}M_\odot$ \citep[][]{paperII}, while it would be $f_{\rm RH}\approx0.3$ in the optimistic model.
Both the models predict $f_{\rm RH}\approx0.6$ around $5\times10^{14}M_\odot<M_{500}<10^{15}M_\odot$, which is the mass range studied in previous observations.
Thus, the future survey on GRHs in low mass clusters ($M_{500}<10^{14.5}M_\odot$) will provide an important clue for the non-thermal contents in GCs.
Note that the difference in the two models stems from the assumption of the slope of the LM relation.
\par

A major revision would be required for the secondary scenario if $f_{\rm RH}\approx0.3$ at $M_{500}\approx 1\times10^{14}M_\odot$ is reported by future observations, since it implies that the radio LM relation is as flat as the optimistic model, while such a model is already in tension with the neutrino upper limit.
The predicted diffuse neutrino intensity can be smaller if the CRP density is smaller than our assumption, i.e., $\epsilon_{\rm CRP}\sim 10^{-13}~{\rm erg/cm^3}$ at the mass of Coma.
Since our model is normalized with the typical luminosity of GRHs, this can be possible when (1) there is a non-negligible amount of primary CREs contributing to the seed population, or when (2) the magnetic field is stronger than our assumption ( $B_0\approx4.7~\mu{\rm G}$) (see also Sect.~\ref{sec:constraint}).
\par

When $f_{\rm RH}\approx0.01$ is in the lower mass range, the radio LM relation should be as steep as the fiducial model, supporting the assumption of TTD acceleration.
Even in this case, the neutrino limit is deep enough to constrain the primary-to-secondary ratio for a given magnetic field.
Thus, the combination of neutrino and radio observations is important to clarify whether seed CREs originated from inelastic $pp$ collisions or not.
\par

Recently, the population of RHs around 140 MHz is being studied with the LOFAR Two-meter sky survey \citep[][]{Shimwell_2017}.
In its second data release \citep[][]{Botteon_2022}, \citet{Cuciti_2023} reported that the LM relation at 150 MHz is consistent with the one extrapolated from 1.4 GHz, assuming the spectral index of $\alpha_{\rm syn} = 1.3$.
The fraction of RHs for $M_{500}>5\times10^{14}M_\odot$ clusters is found to be similar to that at 1.4 GHz \citep[][]{Cassano_2023}.
Although we only focus on the statistics at 1.4 GHz in this work, a detailed comparison between 150 MHz and 1.4 GHz would constrain the parameters fixed in this study, such as $\alpha_{\rm inj}$ and the mass dependence of $B_0$.

\section{Summary and Discussion}
\label{sec:discussion}
Focusing on the so-called secondary scenario, where all of the seed CREs are provided through inelastic $pp$ collisions by CRPs, we estimated the contribution from massive ($M_{500}>3\times10^{14}M_\odot$) clusters to the radio and neutrino backgrounds with the re-acceleration model, which has been often invoked to explain the diffuse radio emission in GCs.
We used the merger tree of dark matter halos to calculate the LFs of the radio and neutrino emissions.
The onset of the re-acceleration is conditioned by the merger mass ratio $\xi$, as the re-acceleration starts only after the merger with $\xi>\xi_{\rm RH}$ \citep[see][for details]{paperII}.
We numerically solved the spectral evolution with the re-acceleration and modeled it with Eq.~(\ref{eq:Luminosity_model}), which is a function of mass and the time since the onset of the re-acceleration.  
Our model parameters are constrained by the observed fraction of GRHs $f_{\rm RH}$ and its mass dependence \citep[][]{paperII}.
\par

In our fiducial model, we adopted $b_{\rm radio} = 3.5$ and $b_\nu = 5.1$ for the slope of the radio and neutrino LM relations. Those relations are reproduced in the re-acceleration model under the assumption of $D_{pp}\propto M^{1/3}$, which is expected in the TTD acceleration in collisionless ICM \citep[e.g.,][]{Brunetti_Lazarian_2011}.
As shown in \citet[][]{paperII}, this model fits well the observed mass trend of the occurrence of GRHs $f_{\rm RH}$ \citep[][]{2021A&A...647A..51C}. 
The radio and neutrino backgrounds are calculated by integrating the LFs with redshift (Eq.~(\ref{eq:isotropic_intensity})).
Due to the relatively steep LM relation, the contribution from massive clusters with $M_{500}>5\times10^{14}M_\odot$ is the largest among clusters.
We tested the optimistic model with shallower slopes of LM relations ($b_{\rm radio} = 1.5$ and $b_\nu = 1.7$). Under this assumption, the effective mass decreases, and the integrated flux of the background emission increases.
Note that this optimistic model tends to predict a larger $f_{\rm RH}$ than the observed one below $8 \times 10^{14} M_\odot$.
\par

We found that the contribution to the ARCADE2 excess is marginal in both models. This conclusion disagrees with the previous estimate by \citet{Fang_Linden:2016dga}, which is based on the turbulent re-acceleration model of \citet{Fujita_2003ApJ...584..190F} with a different set of parameters. 
\citet[][]{Fang_Linden:2016dga} is in tension with the GRH observation, since the predicted synchrotron luminosity is almost two orders of magnitude larger than the typical luminosity of GRHs.

The neutrino intensity is $\sim5$\% of the IceCube level in the fiducial model, while it reaches $\sim20$\% in the optimistic model (Fig.~\ref{fig:IceCube}).
In the fiducial model, the effective neutrino luminosity becomes as large as $E_\nu L_{E_\nu}\approx10^{44}$ erg/s (for all flavors), which is the typical luminosity for the mass of $M_{500} = 1.5\times10^{15}M_\odot$ at $z \approx 0.2$. 
Recent stacking analysis of {\it Planck}-SZ clusters put an upper limit of $\sim5\%$ on the contribution to the diffuse muon-neutrino flux \citep[][]{IceCube_2022_stack}.
The limit is comparable to the expectation in the fiducial model.
\par


We demonstrated that the neutrino upper limit can be used to constrain the central magnetic field strength $B_0$ and the electron-to-proton ratio of primary CRs $f_{\rm ep}$ in the re-acceleration model for GRHs (Fig.~\ref{fig:fep_B}). The current neutrino limit is comparable to the limit for solely the Coma cluster obtained from the gamma-ray upper limit \citet{Brunetti2017}. A neutrino limit as deep as 1\% of the IceCube level would significantly constrain the CRP content in the ICM even in the presence of the re-acceleration.
\par

While nearby massive clusters are not dominant sources for either neutrino or radio backgrounds, our result does not exclude the possibility that GCs with lower masses ($M_{500}<10^{14}M_\odot$) and higher redshifts ($z>1$) are the dominant sources of the neutrino background, as suggested in ``internal source" models \citep[e.g.,][]{Murase2008,Kotera:2009ms,Fang_2018NatPh..14..396F,Hussain:2021dqp}. On the other hand, the recent detection of GRHs in low-mass \citep[][]{Botteon_2021} and high-redshift \citep[][]{DiGennaro_2021NatAs...5..268D} clusters suggests that turbulent re-acceleration also operates in such clusters.
Future studies should follow the long-term evolution of CR distributions, incorporating both the CR injection from AGNs and the re-acceleration.


\acknowledgments
K.N. acknowledges the support by the Forefront Physics and Mathematics Program to Drive Transformation (FoPM). This work is supported by the joint
research program of the Institute for Cosmic Ray Research
(ICRR), the University of Tokyo, and KAKENHI
No. 22K03684 (KA). 
The work of K.M. is supported by the NSF Grant No.~AST-1908689, No.~AST-2108466 and No.~AST-2108467, and KAKENHI No.~20H01901 and No.~20H05852.


\bibliography{Background}{}
\bibliographystyle{aasjournal}

\end{document}